\documentclass[aps,twocolumn,superscriptaddress,showpacs,showkeys,amsmath,amssymb,floatfix]{revtex4}
\usepackage{graphicx}% Include figure files
\usepackage{epsfig}
\usepackage{dcolumn}% Align table columns on decimal point
\usepackage{bm}% bold math
\usepackage{amssymb}
\usepackage{amsmath}
\usepackage{subfigure}
\newcommand{\be}{\begin{equation}}
\newcommand{\ee}{\end{equation}}
\newcommand{\bea}{\begin{eqnarray}}
\newcommand{\eea}{\end{eqnarray}}

\newcommand{\tQ}{{\tilde Q}}

\newcommand{\pro}{\partial}

\newcommand{\ba}{\begin{array}}
\newcommand{\ea}{\end{array}}

\newcommand{\nn}{\nonumber}
\newcommand{\mn}{{\mu\nu}}

\newcommand{\Int}{\displaystyle\int}

\renewcommand{\arraystretch}{1.2}
\makeatletter
\def\@arrayacol{\edef\@preamble{\@preamble \hskip .5\arraycolsep}}
\def\array{\let\@acol\@arrayacol \let\@classz\@arrayclassz
\let\@classiv\@arrayclassiv \let\\\@arraycr\def\@halignto{}\@tabarray}
\makeatother
\begin{document}
\title{Gauge invariant gluon spin operator for spinless non-linear wave solutions}
\bigskip
\author{Bum-Hoon Lee}
\email{bhl@sogang.ac.kr}
\affiliation{Asia Pacific Center of Theoretical Physics,
Pohang, 790-330, Korea}
\affiliation{CQUEST, Sogang University, Seoul 121-742, Korea}
\author{Youngman Kim}
\email{ykim@ibs.re.kr}
\affiliation{Rare Isotope Science Project, Institute for Basic Science,
Daejeon 305-811, Korea}
\author{D.G. Pak}
\email{dmipak@gmail.com}
\affiliation{Asia Pacific Center of Theoretical Physics,
Pohang, 790-330, Korea}
\affiliation{CQUEST, Sogang University, Seoul 121-742, Korea}
\affiliation{Chern Institute of Mathematics, Nankai University,
Tianjin 300071, China}
\author{Takuya Tsukioka}
\email{tsukioka@bukkyo-u.ac.jp}
\affiliation{School of Education, Bukkyo University, Kyoto 603-8301, Japan}
\author{P.M. Zhang}
\email{zhpm@impcas.ac.cn}
\affiliation{Institute of Modern Physics, Chinese Academy of Sciences,
Lanzhou 730000, China}
\begin{abstract}

       We consider non-linear wave type solutions with intrinsic mass scale parameter and zero spin in a pure $SU(2)$ quantum  
       chromodynamics (QCD). A new stationary solution which can be treated as 
       a system of a static Wu-Yang monopole dressed in off-diagonal gluon field is proposed. A remarkable feature
       of such a solution is that it possesses a finite energy density everywhere. All considered non-linear wave type solutions
       have common  features: presence of the mass scale parameter, non-vanishing projection of the color fields along 
       the propagation direction and zero spin. The last property requires revision of the gauge invariant definition
       of the spin density operator which supposed to produce spin one states for the massless vector gluon field. 
       We construct a gauge invariant definition of the classical gluon spin density operator which is unique and 
       Lorentz frame independent.
\end{abstract}
%\vspace{0.3cm}
\pacs{11.15.-q, 14.20.Dh, 12.38.-t}
\keywords{classical solutions in Yang-Mills theory,
gauge invariant decomposition of nucleon momentum, gluon spin operator}
\maketitle
%\vspace{2mm}

\section{Introduction}

There is a common belief that the color confinement and the related mass gap problem
in quantum chromodynamics (QCD) need a consistent non-perturbative quantum theory
for their resolution \cite{colorconft}.
A so-called proton spin crisis \cite{spincr1,spincr2,spincr3} represents another puzzle
which is closely related to non-perturbative dynamics of constituent quarks and gluons.
Besides, recent studies reveal deep conceptual problems in definitions of the
momentum, spin and orbital angular momentum operators of quarks and gluons
(see \cite{review1} for a review and references there in).
In this respect, an important step towards a strict and
non-perturbative theory of QCD is to describe the dynamical content of classical
non-perturbative solutions in a pure $SU(2)$ Yang-Mills theory and establish
their relationship to fundamental observable quantities in QCD
such as vacuum gluon condensate, glueball spectrum and others.

 Non-linear structure and rich topology of the Yang-Mills theory
lead to a wide class of various exact solutions \cite{actor}.
Monopoles and instantons represent the most well-known
topological solutions which have numerous physical implications
(see, for ex., \cite{coleman85,rajaraman}).
Much less is known about non-linear wave type solutions
and especially about their physical meaning.
Non-linear transverse plane wave solutions representing analogues of the electro-magnetic
plane waves were found in 80s by Coleman \cite{coleman}.
The existence of another type of non-linear plane wave solutions with a mass scale parameter
\cite{mat1} is an important manifestation of the conformal symmetry in the
Yang-Mills theory. Various representations of the non-linear plane waves
and some special non-linear spherical wave solutions were considered
in \cite{lahno95,tura,smilga,frasca09,tsap}.
There is a hope that the knowledge of full structure of non-linear wave solutions
in the Yang-Mills theory can provide a novel approach towards
non-perturbative description of quantum chromodynamics.

In the present paper we undertake an attempt to describe a special class
of stationary non-linear wave solutions in a pure $SU(2)$ Yang-Mills theory (or QCD)
which admit intrinsic mass scale parameter. Among such solutions
there are non-linear stationary plane wave solutions \cite{mat1} 
and non-stationary spherically symmetric solutions \cite{p14,p15,p16,p17,p18} which
resemble kink type solitons in two-dimensional coordinate plane $(t,r)$.
It is known that in a pure Yang-Mills theory a stable solitonic
solutions do not exist due to the Coleman theorem \cite{coleman2}. 
This implies that any wave packet solution with a finite total energy 
must decay by radiating its energy to space infinity. It is surprising that a regular stationary monopole like solution 
with a finite energy density everywhere does exist even in  a pure $SU(2)$ QCD
without introducing any additional matter fields. 
We demonstrate the existence of a class of such regular stationary
solutions which represent a system of the Wu-Yang monopole 
interacting to the off-diagonal gluons in a special gauge. 
All considered non-linear propagating solutions and the new proposed stationary monopole like
solutions possess intriguing properties, namely, they admit a mass scale parameter,
non-vanishing longitudinal projections of the color fields and vanished classical spin density. 
This indicates to existence of massive spinless states in the quantum theory. 
Such an idea, that stationary solutions may describe particles (or pseudo-particles)
and even might be related to the vacuum structure in QCD, 
was sounded long time ago \cite{derr,jackiwRMP,jackiw77}. Certainly, to obtain rigorous description
of particle spectrum based on standard QCD as a fundamental theory of strong interaction one has 
to construct a consistent non-perturbative quantization scheme for QCD which remains
an unresolved problem so far. 
The existence of the non-linear wave type solutions
with a non-zero mass and vanished spin manifests inconsistence of the perturbative QCD
which has massless vector gluons in the initial Fock space of physical states.
So even at classical level one should formulate a strict notion of the gluon spin density
operator on the class of massive non-linear wave solutions.  
A consistent definition of the gluon spin and angular momentum operators 
in QCD represents unresolved problem and there is still controversy between different
approaches to this problem \cite{review1}. In particular, there is no unique gauge invariant 
and at the same time explicitly Lorentz frame independent definition for the gluon spin and orbital momentum operators.
In the present paper we show that for the stationary non-linear wave solutions with the mass parameter
it is possible to construct a unique gauge-invariant and Lorentz frame independent gluon spin density operator 
at classical level. 

\section{Non-linear propagating solutions with a mass scale parameter}

In this section we overview briefly the known non-linear propagating wave type solutions which 
possess a mass scale parameter and non-vanishing longitudinal projections of  
color fields \cite{mat1,p14,p15,p16,p17,p18}.

\subsection{Non-linear spinless plane waves}

We consider a class of non-linear wave solutions
admitting mass scale parameters in the
case of a pure $SU(2)$ Yang-Mills theory.
The respective Lagrangian and equations of motion
are as follows
\bea
{\cal L}&=&-\dfrac{1}{4}F_{a\mu\nu}F^{a\mu\nu}, \nn \\
\big(D^\mu F_{\mu\nu}\big)^a &\equiv&
\partial^\mu F_{\mu\nu}^a+g\epsilon^{abc}A^{b\mu} F^c_{\mu\nu} =0,
\label{Lagr0}
\eea
where the field strength is given by
$$
F^a_{\mu\nu}
=\partial_\mu A_\nu^a-\partial_\nu A_\mu^a
+g\epsilon^{abc}A^b_\mu A^c_\nu
$$
with the group structure constants $\epsilon^{abc}$ and the coupling
constant $g$.
The known Coleman non-linear transverse plane wave solutions \cite{coleman}
form a family with six arbitrary functions depending
on light-cone coordinates which correspond to six transverse
propagating massless gluon modes  in agreement with perturbative
description of gluons in QCD.
An important feature of the non-Abelian gauge theory is that
it admits a three-parametric family of non-linear massive wave solutions
with a field strength having non-vanishing longitudinal projections
along the wave vector \cite{lahno95,tura,smilga,frasca09}.
Such plane waves can be interpreted as non-perturbative
longitudinal modes of gluon.  The solutions can be reproduced
by using a simple ansatz
\bea
A_i^a(x)&=&\delta_i^a \phi_i(u),~~~~~A_0^3(x)=\beta \phi_3(u), \label{ansf1234}
\eea
where $u\equiv k_0 t+ k_3 z$ and $\beta=v/c$ is a kinematic
variable proportional to wave velocity ``$v$'' in units of the
speed of light
``$c$''.
The non-vanishing field strength components are the following: 
\renewcommand{\arraystretch}{1.4}
\begin{equation}
\begin{array}{rcl}
F^1_{10}&=&-\pro_0 \phi_1,~~~~~~~~F^1_{20}=g\beta \phi_2\phi_3, \\
F^2_{10}&=&-g\beta \phi_1\phi_3,~~~~F^2_{20}=-\pro_0 \phi_2, \\
F^3_{30}&=&-(1-\beta^2)\pro_0\phi_3, \\
F^1_{13}&=& -\beta \pro_0 \phi_1,~~~~~~F^2_{13}=-g\phi_1 \phi_3, \\
F^1_{23}&=&g \phi_2 \phi_3 ,~~~~~~~~F^2_{23}=-\beta\pro_0 \phi_2, \\
F^3_{12}&=& g\phi_1 \phi_2.
\end{array}
\label{fieldstr}
\end{equation}

\noindent
One can verify that ansatz (\ref{ansf1234}) leads to
non-vanishing electric and magnetic longitudinal projections.
By direct substitution of the ansatz into the Yang-Mills equations
and imposing a constraint $k_3=\beta k_0$ one can
reduce all equations of motion to three ordinary differential equations (ODE)
\bea
&& k^2 \dfrac{{\rm d}^2\phi_1}{{\rm d}u^2}+g^2 \phi_1
\Big(\phi_2^2+(1-\beta^2)\phi_3^2\Big)=0, \nn \\
&& k^2\dfrac{{\rm d}^2\phi_2}{{\rm d}u^2}
+g^2 \phi_2 \Big(\phi_1^2+(1-\beta^2)\phi_3^2\Big)=0, \label{eqbeta} \\
&& k^2\dfrac{{\rm d}^2\phi_3}{{\rm d}u^2} +g^2 \phi_3
\Big(\phi_1^2+\phi_2^2\Big)=0,  \nn
\eea
where  $k^2 \equiv k_0^2-k_3^2$. Note that
the constraint  $k_3=\beta k_0$ provides
 a  Lorenz gauge condition $\pro^\mu A_\mu^a=0$ for the
gauge potential.
In the rest frame, $\beta=0$,
the equations (\ref{eqbeta}) describe
a classical mechanical system of three anharmonic oscillators
with the following Hamiltonian
\bea
H&=&\dfrac{1}{2}\big(\dot \phi_1^2+\dot \phi_2^2+\dot \phi_3^2\big)
+\dfrac{g^2}{2} \big(\phi_1^2\phi_2^2+\phi_2^2\phi_3^2+\phi_3^2\phi_1^2\big).
\quad \label{Ham}
\eea
In general, such a system represents a
non-integrable problem \cite{mat-savv}.
It has been proved as well that
the corresponding quantum mechanical system possesses a
pure discrete energy spectrum despite the presence of zero energy valleys
in the Hamiltonian \cite{simon83}.

Let us consider two special cases when the system of equations
(\ref{eqbeta}) becomes integrable.
The first case, (I), corresponds to a constraint
 $\phi_1=\phi_2\equiv \phi$, $\phi_3=0$, and, the second case, (II),
is determined by setting $\phi_1=\phi_2=\phi_3\equiv \phi$ and imposing an additional
condition $\beta=0$.
Corresponding equations are the following
\renewcommand{\arraystretch}{2.5}
\be
\begin{array}{rcl}
{\rm (I)}:&~&\displaystyle
k^2 \frac{{\rm d}^2\phi(u)}{{\rm d}u^2}+g^2 \phi^3(u)=0, \\
{\rm (II)}:&~&\displaystyle
k^2 \frac{{\rm d}^2\phi(u)}{{\rm d}u^2}+2 g^2
 \phi^3(u)=0.
\end{array}
\label{eq1}
\ee
\noindent
Various representations of these equations
have been obtained in $SU(2)$ Yang-Mills theory
by using different methods \cite{tura,lahno95,smilga,frasca09,mat-savv}.
Solutions to the equation (\ref{eq1}) are given by the Jacobi elliptic
function
\renewcommand{\arraystretch}{2.0}
\be
\begin{array}{rcl}
\phi_{\rm I}(u)&=&\dfrac{\sqrt 2 \mu_1}{\sqrt g}\, {\rm sn}
\Big[\dfrac{\mu_1}{k} \sqrt g (u+u_{01}),-1\Big], \\
\phi_{\rm II}(u)&=&\dfrac{\mu_2}{\sqrt {g}}\, {\rm  sn}
\Big[\dfrac{\mu_2}{k} \sqrt g (u+u_{02}),-1\Big],
\end{array}
\label{phi0}
\ee
\noindent
where $\mu_i, u_{0i}$ are integration constants. One can set $u_{0i}=0$ since $u_{0i}$
is the parameter corresponding to translation invariance.
The argument of the elliptic function can be re-written
in a Lorentz invariant form as $p^\mu x_\mu$
which implies a dispersion relation
$p^2=\mu^2 g$.
One should stress that two solutions (\ref{phi0}) are gauge non-equivalent
and have different implementations in the mathematical structure of the Yang-Mills theory.
  
\subsection{Spherically symmetric non-stationary finite energy solutions}

   In this subsection we overview briefly known spherically-symmetric non-stationary solutions
\cite{p14,p15,p16,p17,p18}. We consider a restricted class
of such solutions which have a spherically symmetric
initial shape in the rest frame, i.e.,  the total linear
momentum is supposed to be zero.
A class of finite energy solutions is described by the following ansatz
for the gauge potential in spherical coordinates
 $(r,\theta,\varphi)$
with one unconstrained function $\psi(t, r)$ \cite{p14,p15,p16,p17,p18} 
%Apr16Jan19spherWPacket-2chck2.nb
\bea
A_m^a= -\epsilon^{abc}\hat n^b \pro_m \hat n^c (1+\psi(t,r)), \label{wuyang}
\eea
where $\hat n=\vec r/r$. It is clear, that the ansatz (\ref{wuyang})
describes a generalized time dependent Wu-Yang monopole configuration.
A known static Wu-Yang monopole corresponds to a special case $\psi(t,r)=0$,
and a trivial pure gauge vacuum configuration is described by $\psi(t,r)= \pm 1$.
Direct substitution of the ansatz into the equations of motion
leads to one non-trivial independent partial differential equation
\bea
\pro_{t}^2\psi-\pro_{r}^2\psi+\dfrac{1}{r^2}\psi(g^2\psi^2-1)=0. \label{eqspher}
\eea

The equation (\ref{eqspher}) represents a second order hyperbolic equation
which admits a wide class of wave solutions determined uniquely
by given initial conditions. We give one example of finite energy solutions
which looks like a kink in two-dimensional coordinate plane $(t,r)$
(others can be found in \cite{p14,p15,p16,p17,p18}).
To find a numeric solution one chooses an initial profile function $\psi(t=0,r)$
in such a manner that the magnetic field vanishes at space
infinity $\psi(t=0,r)=1-r^2 {\rm e}^{-r^2}$.
This profile function describes a monopole like configuration
with a maximal magnetic charge at a finite distance from
the center, and a vanishing total magnetic charge at large distances.
A corresponding numeric solution demonstrates
a soliton structure in the effective 1+1 space-time $(t, r)$;
the solution $\psi(t, r)$ represents a lump
in the coordinate plane $(t, r)$, FIG. \ref{Fig5}, which moves
in radial direction to space infinity with the light speed.
\begin{figure}[htp]
\centering
\includegraphics[width=70mm,height=55mm, bb=0 0 541 356]{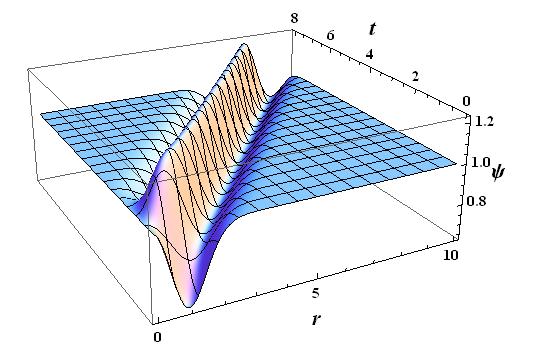}
\caption[fig5]{A kink type solution
with initial conditions $\psi(0, r)=1-r^2 {\rm e}^{-r^2},
~\pro_t \psi(0, r)=0$.}\label{Fig5}
\end{figure}
Note that one finds a similar behavior of the magnetic field for the
system of localized monopoles and antimonopoles in the Weinberg-Salam
model \cite{pakIJMP15}.

We conclude that a pure QCD admits a wide class
of non-linear wave solutions. 
All solutions carry zero spin and the non-linear plane waves
has the dispersion relation with the mass parameter.
Taking into account that a general solution has
additional three parameters corresponding to orientation
in the internal space of the group $SU(2)$,
one can interpret the non-linear pane waves as
three longitudinal dynamic degrees of freedom
in addition to six transverse dynamic degrees
of freedom represented by Coleman non-Abelian
plane waves \cite{coleman}.
Note that transverse non-linear waves remain massless,
so the mass appears only from the solutions containing
longitudinal projections of the field strength.
This is contrary to other approaches like the models of massive gluodynamics
where the gluon mass is introduced either through spontaneous symmetry breaking
or by adding explicit mass terms \cite{math}.
A principal advantage of our treatment is that the color gauge symmetry
remains unbroken, so the theory is still renormalizable in a standard perturbative
quantum field theory.

\section{Wu-Yang monopole dressed in off-diagonal gluon field}

Let us rewrite the ansatz  (\ref{wuyang})
by performing an appropriate $SU(2)$ gauge transformation
 in a so-called Abelian gauge \cite{choprd80}
as follows
\bea
A_\theta^2&=&\psi (t, r), \nn \\
A_\varphi^1&=& -\psi(t, r) \sin \theta, \label{spherwav} \\
A_\varphi^3&=&\dfrac{1}{g} \cos \theta.  \nn
\eea
All other components of the gauge potential are identically zero.
The expression for the Abelian component $A_\varphi^3$  contains coordinate singularity
along the $z$-axis, so such a gauge represents a singular gauge.
One should stress, that final  gauge-invariant quantities (like the Lagrangian and energy density)
are regular and do not depend on a chosen gauge.
We prefer the ansatz written in the Abelian gauge \cite{choprd80}, 
(\ref{spherwav}), since such notation is more suitable for description of stationary
monopole solutions in $SU(N)$ Yang-Mills theory and multimonopole 
configurations. Besides, due to gauge invariant decomposition of the gauge potential \cite{choprd80}
one can treat the Abelian gauge potential $A_\varphi^3$ as a static Wu-Yang monopole field
and the function $\psi(t,r)$ as a dynamic degree of freedom describing the off-diagonal gluon.

We are interested in regular stationary wave solutions
to equation (\ref{eqspher}). Let us write down 
the energy functional corresponding to the ansatz (\ref{spherwav}), 
\bea
E&=&\int\!{\rm d}r\, {\rm d}\theta\, {\rm d}\varphi\,
\sin \theta\Big ((\pro_t \psi)^2+(\pro_r \psi)^2
\nn
\\
&&\hspace*{27mm}
+\dfrac{1}{2g^2r^2}(g^2 \psi^2-1)^2 \Big ) \nn \\
  &\equiv& 4 \pi \int\! {\rm d}r \, {\cal E}(t, r), \label{entotspher}
\eea
where ${\cal E}(t, r)$ is an effective energy density
defined on (1+1)-dimensional half-plane $(r\geq 0, \ 0<t<\infty)$.
The effective energy density can be treated as
a Hamiltonian of two-dimensional $\lambda \phi^4$ theory
with a radial dependent coupling $\lambda \equiv 1/(2g^2r^2)$.

The field strength components
contain the following non-vanishing projections
of the color magnetic and electric field
\bea
F_{r\theta}^2&=&\pro_r \psi,~~~~~~~F_{r\varphi}^1=-\pro_r \psi \sin\theta, \nn \\
F_{\theta\varphi}^3&=&\dfrac{1}{g}(g^2\psi^2-1)\sin \theta,  \label{Ftf} \\
F_{t\theta}^2&=&\pro_t \psi, ~~~~~~~F_{t\varphi}^1=-\pro_t\psi\sin\theta. \nn
\eea
The radial magnetic field component $F_{\theta\varphi}^3$
generates a non-zero magnetic flux through a sphere with a center
at the origin, $r=0$.
So that, the color magnetic charge of the monopole depends on time and distance from the center. 
Note that various static generalized Wu-Yang monopoles have been considered before, 
however, all of them have singularities in agreement with the
Derrick's theorem \cite{derr}.

The equation (\ref{eqspher}) admits a local non-static solution near the origin
which removes the singularity of the monopole at the center 
\bea
\psi(t,r)&=&\dfrac{1}{g}+\sum_{n=1} c_{2 n}(t) r^{2 n},  \nn \\
c_4(t)&=& \dfrac{1}{10} \big (3 g c_2^2(t) +c_2''(t)\big ),   \nn \\
 c_6(t)&=&\frac{1}{28}\big(c_4''(t)+6gc_2(t)c_4(t)+g^2c_2^3(t)\big), \nn \\
&\vdots&       \label{locsol}
\eea
where the coefficient functions $c_{2n}(t)$ ($n\geq 2$) are determined in 
terms of one arbitrary function $c_2(t)$. 
 In asymptotic region, $r\simeq \infty$,
the non-linear 
equation (\ref{eqspher}) reduces to a free D'Alembert
equation which has a standing spherical wave solution
\bea
\psi(t, r) &\simeq& a_0+A_0 \cos (M r) \sin(M t)
+{\mathcal O}\Big(\dfrac{1}{r}\Big), \label{asymsol}
\eea
where $a_0, A_0$ are integration constants,
the mass scale $M$ appears due to scaling invariance 
in the theory under dilatations $r \rightarrow Mr, t\rightarrow Mt $. 
One should stress, that the asymptotic solution represents a standing spherical wave only 
in the leading order and our solution can not be treated as a superposition 
of out-going and in-going spherical waves due to absence of superposition principle in the
non-linear theory. Besides, as we will see below, the parameters $a_0, ~A_0$ are not independent
free parameters. Moreover, all known before non-linear spherical wave solutions in the Yang-Mills theory
are singular. Note that the series expansion for the local solution 
starts from the factor $1/g^2$ which reflects interrelationship of the solution
with the non-perturbative topological origin of the Wu-Yang monopole. In particular, 
the presence of such a term cancels the singularity of the Wu-Yang monopole.
The second term in the series expansion, $c_2(t) r^2$, contains quadratic dependence on the radial
coordinate. This provides finiteness of the energy density at the origin. 

\begin{figure}[h!]
\centering
\includegraphics[width=70mm,height=52mm,bb=0 0 562 377]{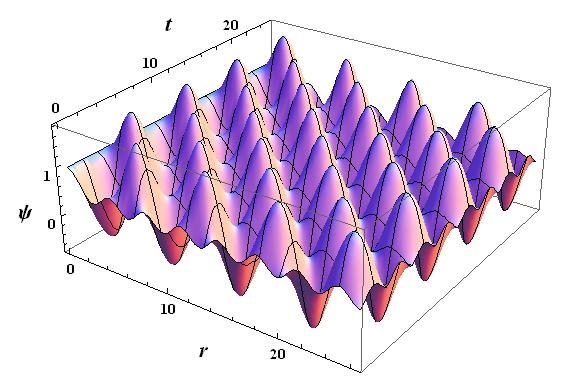}
\caption[fig1]{Stationary monopole solution;
$(0\leq r,t \leq L)$, ~$L=8 \pi$, $c_0=-0.041$,~$c_1=-0.523,~g=1$
.}\label{Fig1}
\end{figure}

The equation (\ref{eqspher}) represents a hyperbolic partial differential equation
and admits a correct Cauchy problem setup with arbitrary
initial conditions for the function $\psi(t,r)$  at the initial time $t=0$.
For instance, one can choose any initial profile function periodic along the
radial direction. However, such a solution will not be stationary in general.
To find a stationary solution we will solve a Cauchy problem imposing
initial conditions with a periodic in time initial conditions at the origin $r=0$.
To solve numerically such problem one chooses a rectangular numeric 
domain $(L_0 \leq r \leq L,~ 0\leq t \leq L)$. Since at $r=0$ one has coordinate
singularity which implies the stiffness numeric problem we introduce a small number
$L_0$ and then check convergence of the numeric solution in the limit $L_0 \rightarrow 0$.
We use the local solution  (\ref{locsol}) 
to impose initial Dirichlet conditions along the boundary $r=L_0$.
 The initial profile function $c_2(t)$ can be chosen arbitrarily as any regular
periodic function, we set it for simplicity in terms of ordinary sine function
\begin{equation}
 c_2(t)=c_0+c_1 \sin (M t),  \label{locsol10}
\end{equation}
where $c_0, \, c_1$ are numeric constants. 
Note that only one of two parameters in the local solution (\ref{locsol10})
is free, the other is fixed by the requirement that a
numeric solution matches the asymptotic solution (\ref{asymsol}).
Dimensional analysis implies that the energy of the solutions
is proportional to $M$ so that the energy vanishes in the limit $M\rightarrow 0$.
This might cause some doubts on existence of the solution.
However, one should stress that standard arguments
on existence of static solitonic solutions based on the Derrick's theorem
\cite{derr} are not applicable to the case of stationary solutions 
which satisfy the extremum principle of the classical action, 
not the energy functional. In  the case of Yang-Mills theory the action is conformal invariant
and its first variational derivative with respect to the scale parameter $M$ vanishes
identically. So the parameter $M$ determines the scale of the space-time coordinates
and can be set to one without loss of generality.
With this one can solve numerically the equation (\ref{eqspher}), and
the corresponding solution is depicted in Fig.\ref{Fig1}.
Note that one has a stiffness numeric problem near the origin, so that  
we have checked the regular structure and convergence of the numeric solution 
in close vicinity of the origin up to $L_0=10^{-6}$ while keeping the 
radial size of the numeric domain of order $L=64 \pi$. 

A general stationary monopole 
solution can be classified by two of three parameters $a_0, A_0, M$ 
characterizing the asymptotic behavior of the solution (\ref{asymsol}). 
The mass scale parameter $M$ takes arbitrary values
whereas the mean value $a_0$ and amplitude $A_0$ parameters
are constrained. Numerical analysis implies the following dependence
of the amplitude $A_0$ of the monopole solution on its mean value parameter
$a_0$, Fig. 3. The dependence of the amplitude $A_0$ on the mean value $a_0$ of the monopole
solution is important in study of the quantum stability of the vacuum gluon condensation 
based on using the classical stationary monopole solution \cite{p2}.
\begin{figure}[htp]
\centering
\includegraphics[width=70mm,height=55mm,bb=0 0 324 213]{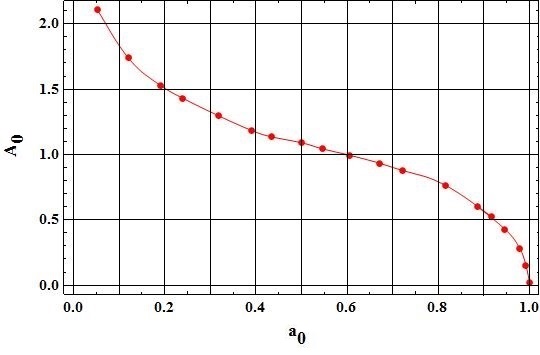}
\caption[fig8]{Dependence of the amplitude $A_0$ of the monopole solution on the 
mean value $a_0$,~ $M=1$.}\label{Fig8}
\end{figure}

Note that in the case of vanishing function $\psi(t, r)$ one has still a non-trivial solution
for the Wu-Yang monopole which has a singularity at the origin $r=0$.
The monopole is defined by the Abelian gluon field $A_\varphi^3=\cos \theta$
and the function $\psi(t, r)$ describes off-diagonal gluons.
The obtained solution can be treated as a Wu-Yang monopole dressed in a
spherical standing wave made of off-diagonal gluons. It is surprising that
the spherical standing wave regularizes the singularity of the Wu-Yang monopole
resulting in a finite energy density at the origin $r=0$. Another interesting feature
of the solution is that the standing wave does not screen completely the color monopole
charge at large distances.  One can find that in the asymptotic region the function $\psi(t,r)$
oscillates around the value $b_0\simeq 0.65$.
So the radial component of the color magnetic field
$F_{\theta \varphi}^1$ has a non-zero averaged value
which provides a non-vanishing total color magnetic charge.

One can easily generalize the above consideration of $SU(2)$ stationary
monopole solutions to the case of $SU(N)$ Yang-Mills theory.
For the case of a pure $SU(3)$ QCD one has the following ansatz
for generalized Wu-Yang monopole solution corresponding to 
color magnetic charge two
 \bea
&A_\theta^2=\psi_1 (t, r),~~~~~~~~~~~~~&   A_\theta^5=\psi_2 (t, r), \nn \\
&A_\varphi^1= -\psi_1(t, r) \sin \theta, ~~~~~& A_\varphi^4= \psi_2(t, r) \sin \theta, \qquad\label{spherwav3} \\
&A_\varphi^3= \dfrac{1}{g} \cos \theta, ~~~~~~~~~~~~~~&
A_\varphi^8=-\dfrac{\sqrt 3}{g} \cos \theta, \nn
\eea
where the Abelian gauge potentials
$A_{\varphi}^{(3,8)}$, corresponding to generators of the Cartan subalgebra of $SU(3)$,
describe a static Wu-Yang monopole with a total color magnetic charge two, 
$g_m^{\rm tot}=2$, \cite{choprl80}. The functions $\psi_{1}(t,r)$ and $\psi_{2}(t,r)$ 
correspond to dynamic degrees of freedom of the off-diagonal components of the gluon field.
One can verify that substitution of the ansatz  (\ref{spherwav3}) into the equations of motion
of the pure $SU(3)$ QCD implies two independent partial differential equations for two functions 
$\psi_{1,2}$
\be
\begin{array}{rcl}
&&\pro^2_{t} \psi_1 -\pro^2_r \psi_1+\dfrac{g^2}{2 r^2} \psi_1
\Big(2 \psi_1^2-\psi_2^2-\dfrac{2}{g^2}\Big)=0, \\
&&\pro^2_{t} \psi_2 -\pro^2_r \psi_2+\dfrac{g^2}{2 r^2} \psi_2
\Big(2
 \psi_2^2-\psi_1^2-\dfrac{2}{g^2}\Big)=0.
\end{array}
\label{eqsu3}
\ee
In a special case, $\psi_1(t,r)=\psi_2(t,r)\equiv \psi(t,r)$, the equations (\ref{eqsu3})
reduce to one differential equation
 \bea
&&\pro^2_{t} \psi -\pro^2_r \psi+\dfrac{g^2}{2 r^2} \psi
\Big(\psi^2-\dfrac{2}{g^2}\Big)=0. 
\eea
By simple rescaling $\psi(t,r) \rightarrow \sqrt 2 \psi(t,r)$ 
the last equation is transformed to the equation $(\ref{eqspher})$
for $SU(2)$ monopole solution. 

Existence of stationary solutions with a magnetic charge and a finite energy density
is unexpected in a pure Yang-Mills theory since there is no such an analogue
in linear field models including the Maxwell type gauge field. Remind, that
a known 't Hooft-Polyakov monopole solution includes external Higgs scalar
fields which regularize the monopole singularity at the origin.
Besides, one can verify that the standard electroweak theory does not admit
such regular stationary monopole solutions.
So that, the quantum chromodynamics is a unique theory
among the currently known theoretical field models of fundamental interactions
which possesses such a surprising feature.

\section{Stability analysis}

       Stability of static localized solitonic solutions can be verified by checking whether the second variation 
         of the energy with respect to small deformations becomes negative.
         If solution possesses a specific space symmetry one should
        take into account perturbations of most general field configuration as well. 
        We will show that the stationary spherically symmetric monopole solution 
        considered in the previous section
        is unstable against small axially-symmetric perturbations $Q_\mu^a(r,\theta,t)$ 
        around the classical solution determined by the ansatz
        (\ref{spherwav}).
%\bea
%A_\mu^a=A^{cl~a}_\mu+Q_\mu^a.
%\eea
In the case of stationary solutions with infinite total energy one can apply a similar approach 
to stability problem as in the case of solitons. We study an eigenvalue spectrum of the operator $\hat K_{\mu\nu}^{ab}$
defined by means of second variational derivatives of
the classical action $S[A]$
\bea
&&\hat K_{\mu\nu}^{ab} Q_\nu^a=\lambda Q_\mu^a,   \label{schreq} \\
&&\hat K_{\mu\nu}^{ab} \equiv \dfrac{\delta^2}{\delta Q_\mu^a \delta Q_\nu^b}  {S[A_{cl}+Q]}.  \label{operK}
\eea
%Euler-Lagrange equations determine
%field configurations which provide extremal values of the action, either local minimum
%or maximum or a saddle point which implies instability.

It is instructive to show that despite on infinite total energy and asymptotic oscillating behavior 
of the Lagrangian corresponding to the stationary monopole solution
the classical action is finite and well-defined.
One can check that non-linear plane wave and stationary monopole solutions
realize an absolute maximum of the classical action. 
Let us verify this for the stationary monopole solution.
A solution for the stationary monopole can be represented in terms of Fourier series 
as follows
\bea
\psi(t,r)=C_0(r)+\sum_{n=1,2,3,...} P_n(r) \cos(n r) \sin (n t), \label{series}
\eea
where one has only cosine functions in the decomposition due to the structure of the local solution,
(\ref{locsol}), and the radial functions  $C_0(r), P_n(r)$  satisfy
 the boundary condition at the origin, $C_0(0)=1$, $P_n(0)=0$, and asymptotic solution (\ref{asymsol}), i.e., 
 $P_n(r=\infty)=A_0$.
 
Let us consider for simplicity a leading mode approximation keeping only the first term in the series decomposition
(\ref{series}).
The functions $C_0(r), P_1(r)$ can be decomposed in series in degrees of orthogonal polynomials. For our purpose to
demonstrate that the variational method can be applied successfully to the classical action we choose a simple variational trial function for the monopole solution 
\bea
\psi(t,r)&=&1-\dfrac{(1-a_0) r^2}{1+b_0 r^2}\nn \\
                                &+&A_0 (1-e^{-d_0 r^2}) \cos (M r) \sin(M t), \label{interpolf}
\eea
where the conformal scale factor $M$ will be set to one in numeric calculations,
$a_0,b_0,A_0, d_0$ are fitting number parameters.
Note that in the case of the stationary monopole solution the convergence of the series (\ref{series})
is very fast and the expression (\ref{interpolf}) provides a good interpolation function
to the exact numeric solution. 
Substituting the trial function (\ref{interpolf}) into the action one can perform integration over the time in the 
interval $(0\leq t \leq 2 \pi)$,
integration over the spherical angles $(\theta, \varphi)$ is trivial and leads to a number factor $4\pi$.
The obtained effective action $A^{eff}$ is defined in one-dimensional space $(0\leq r \leq \infty)$
and contains a divergent term $A^{div}$ which originates 
from the kinetic terms in the original Lagrangian
\bea
A^{div}
% &=&  4 \pi^2 \int dr A_0 (\cos^2 r-\sin^2 r)  \nn \\
           &=& 4 \pi^2 \int dr A_0 \cos(2 r).
\eea
A simple regularization by introducing an exponential factor 
$e^{-\varepsilon r}$ with an infinitesimally small number $\varepsilon$ leads to a vanished
value of the divergent term.
%Note that in the general 
%non-spherically symmetric case there are also logarithmic and linear divergent terms 
%with oscillating behavior. A detailed analysis shows that all divergent terms
%produce finite results after proper regularization and integration over the radial coordinate.
The remaining part of the effective action is well-defined and integration over the radial coordinate 
can be easily performed numerically. With this, the action represents a regular function 
of the trial parameters $(a_0, A_0, b_0, d_0)$. The first two parameters represent a mean value and amplitude
of the monopole solution in the asymptotic region, the parameters $b_0, d_0$ describe 
matching profile of the local and asymptotic solutions along the radial direction. 
One can fix one of two parameters $a_0, A_0$ since only one of them is independent, (see Fig. 3), 
all other parameters are found by variational procedure which finds an extremum the action. 
The numeric results for a given asymptotic amplitude $A_0=0.56$ 
show that the classical action has an absolute maximum, $S^{max}=-0.168...$,
with the following trial parameters values
%Feb15OneMonstabilityCHCK2.nb
\bea
a_0&=&0.9982..., \nn \\
b_0&=&0.02588..., \nn \\
d_0&=&0.1786...\, .
\eea
The results are in a good qualitative agreement with the numeric results for the 
stationary monopole solution presented in Figs. 2,3.

Now we consider the stability of the monopole solution under axially-symmetric 
perturbations. We consider a small perturbation $Q(r,\theta,t)$ 
around the Abelian gauge potential $A_\varphi^3$ which describes a static Wu-Yang monopole
within the framework of the ansatz  (\ref{spherwav})
\bea
 A_\varphi^3&=& \cos \theta+Q(r,\theta)\cos (M t).  \label{flucfuncs}
 \eea
We constrain our study by consideration of perturbations with time dependent factor $\cos (M t)$ 
 with the same frequency $M$ as one in the monopole solution. 
  It is suitable to pass to standard notations in spherical coordinates defining
 the perturbation field $\tilde Q(r,\theta)$ of mass dimension
 \bea
 \tilde Q(r,\theta)&=& \dfrac{1}{r \sin\theta} Q(r,\theta).
 \eea
For brevity of notations we denote the interpolation function $\psi(t,r)$, (\ref{interpolf}), 
as follows
\bea
\psi(t,r)&=&1+P_0(r) \cos (M t).
%P_0(r)&=& A_0 \cos r.
\eea  
In the case of small perturbations it is enough to keep only terms quadratic in $Q(r,\theta)$ 
 in the classical action. Substituting the perturbed potential $A_\varphi^3$, (\ref{flucfuncs}),
 into the classical action and performing
 integration over $(t,\varphi)$ one obtains the following 
 quadratic form for the operator $\hat K_{\mu\nu}^{ab}$,  (\ref{operK}),
 %Feb16OneMonstabilityAxialFlucts.nb
 \bea
v{\cal L}^{(2)}&=& 2 \pi^2 \Big  [ -r^2 \sin^2 \theta (\pro_r \tilde Q)^2-\sin\theta (\pro_\theta \tQ)^2   \nn \\
&-&
2 \tQ(\cos \theta \pro_\theta \tQ+r \sin \theta \pro_r \tQ)- \Big (\csc\theta+\nn \\
 &&(1-M^2 r^2 +\dfrac{3}{4} P_0^2\Big) \sin \theta \tQ^2
\Big ] ,
\eea
  where $v\equiv r^2 \sin \theta$ is an integration volume in spherical coordinates.
  Varying the last expression with respect to $\tilde Q$ one can write down explicitly the eigenvalue equation
  (\ref{schreq})  for possible unstable modes  
  \bea
&& -r^2 \sin\theta  \Big [-\pro_r\pro_r-\dfrac{2}{r}\pro_r-\dfrac{1}{r}(\pro_\theta\pro_\theta +\cot \theta  )
+V \Big ] \tQ \nn \\ 
 && 
= \lambda \tQ,  \label{eigeq} \\
&& V \equiv -M^2 + \dfrac{1}{r^2} \Big ( 1+\dfrac{1}{\sin^2\theta} +\dfrac{3}{4} P_0(r) \Big). 
\eea
The equation resembles a Schrodinger type equation (up to opposite sign on the left hand side) 
with a quantum mechanical potential $V$. Simple consideration shows that
due to presence of the factor $-M^2$ the potential $V$ is negative at large distance.
This may cause appearance of negative eigenvalues 
for such a ``Schrodinger'' equation, or, equivalently, positive values for $\lambda$
in the original eigenvalue equation (\ref{eigeq}). Corresponding eigenfunctions 
represent unstable perturbation modes which will increase the value of the non-perturbed action. 
Therefore, the spherical monopole solution will be unstable under axially-symmetric field perturbations. 
Note that in the case of spherically symmetric perturbations around the ansatz (\ref{spherwav})
a similar term $-M^2$ in the potential $V$ does not imply negative eigenvalues. A detailed 
analysis shows that due to scaling invariance there are only zero perturbation modes corresponding
to zero eigenvalues. 

To solve numerically the equation (\ref{eigeq}) we impose the following boundary conditions
for the perturbation field $\tQ$
A numeric solution corresponding to a positive eigenvalue closest to zero 
is presented in  Fig. 4a. The unstable mode $\tQ_1(r,\theta)$ 
has oscillating behavior with an amplitude decreasing in radial direction as $\dfrac{1}{r^2}$.
\bea
\tQ(\infty,\theta)&=& 0 , ~~~~~~~\tQ(r, 0)= \tQ(r,\pi).
\eea
%\begin{widetext}
%Feb21OneMonStCORR-CHCK-u3only.nb
%Feb21OneMonU3L8pi_w=1regextrem2.mph
 \begin{figure}[h!]
\centering
\subfigure[~]{\includegraphics[width=80mm,height=50mm]{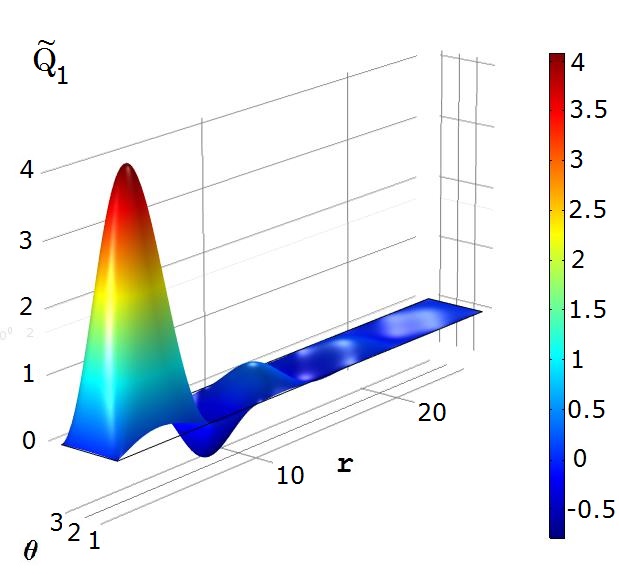}}
\hfill
\subfigure[~]{\includegraphics[width=80mm,height=60mm]{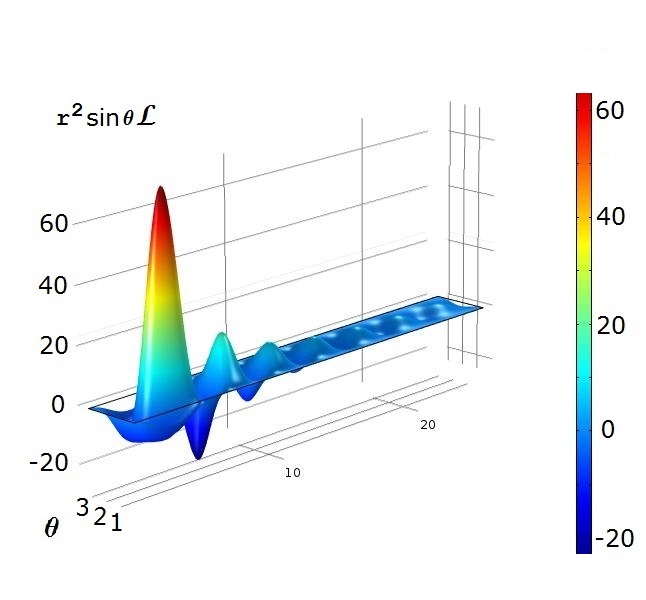}}
\caption[fig4]{(a) Perturbation mode $\tQ_1(r,\theta)$ of even parity corresponding to the eigenvalue $\lambda_1=+1.498$; %The normalization of $\tQ_1$ can be chosen arbitrary due to linearity of the eigenvalue equation (\ref{eigeq});
(b) an integral density
$r^2 \sin\theta {\cal L}$ for the second variation of the classical action.}\label{Fig4}
\end{figure}
\begin{figure}[h!]
\centering
\subfigure[~]{\includegraphics[width=80mm,height=55mm]{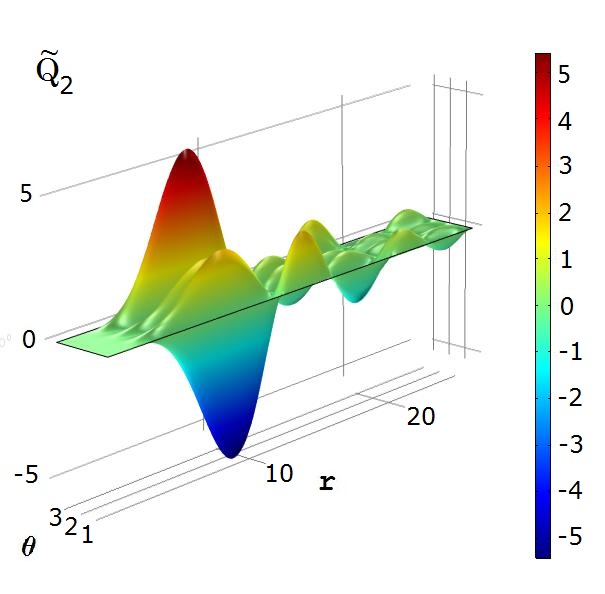}}
\hfill
\subfigure[~]{\includegraphics[width=80mm,height=55mm]{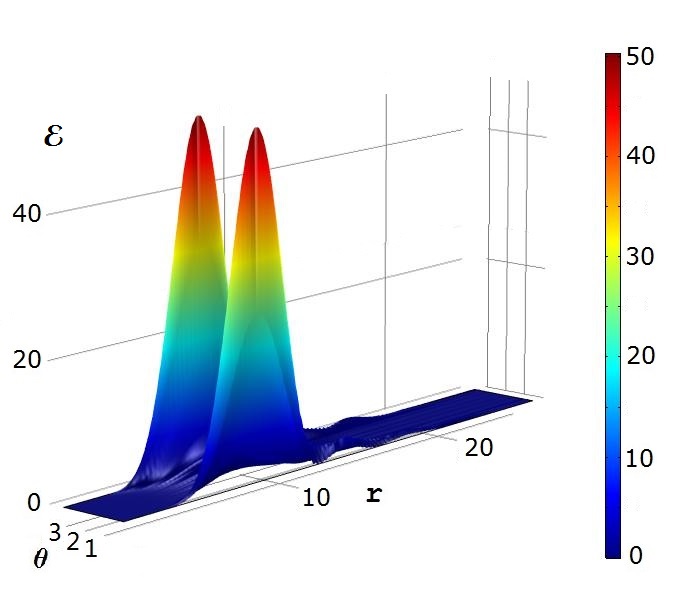}}
\caption[fig5]{(a) Perturbation mode $\tQ_2(r,\theta)$ of odd parity corresponding to the eigenvalue $\lambda_2=+2.320$;
(b) a respective energy density $\cal{E}$.}\label{Fig5}
\end{figure}
Note that since the eigenvalue equation (\ref{eigeq}) is linear
one can normalize the function $\tQ_1(r,\theta)$ to any small number. We keep non-renormalized
solutions $\tQ(r,\theta)$ in our numeric results.
Numeric estimate of the second variation of the classical action
confirms that it is positive. So, the perturbations increase the extremum value of the action 
showing instability of the spherically symmetric monopole.

Finite energy perturbation modes can be classified into  
two type field configurations with respect to reflection symmetry 
$\theta \rightarrow \pi-\theta$. The first mode $\tQ_1$ keeps
its shape under the reflection and has even parity.
An unstable mode 
corresponding to the next positive eigenvalue is depicted in Fig. 5a,
and it represents odd parity field configuration.
Unstable modes with larger positive eigenvalues correspond to presence of
higher Fourier modes in polar angle $(\theta)$.
All regular unstable modes with appropriate asymptotic behavior 
lead to a finite total energy and finite second variation of the classical action.
In particular, the energy density ${\cal E}_1$ corresponding to the mode $\tQ_1$ has integrable singularities 
at two points on $Z$-axis, $(r\simeq \pm  0.72, \theta=0,\pi)$, and the energy density 
 ${\cal E}_2$ corresponding to the mode $\tQ_2$ 
has a finite maximum along two rings with centers located on the $Z$-axis
at finite distance from the origin, Fig. 5b.

Remind that $\tQ(r,\theta)$ is a perturbation around the Abelian gauge field $A_\varphi^3(r,\theta)=\cos\theta$ which describes a static Wu-Yang monopole. 
The presence of two types of unstable modes with energy density maximums located near the $Z$-axis at finite
distance from the origin indicates to possible existence of two types of axially symmetric solutions
which might correspond to monopole-monopole and 
monopole-antimonopole pairs.  

In general, in non-linear theories 
the stability of stationary solutions represents a non-trivial problem.
We perform analysis of the stability in linear approximation 
for axially-symmetric perturbation around the Abelian gauge potential $A_\varphi^3$. 
In this stage we may conclude
that spherically symmetric monopole is unstable under small axially-symmetric perturbations.

\section{Gluon spin density operator}

The classical non-linear solutions considered in the present paper have
common features: all of them have a mass scale parameter, longitudinal projections of color fields
and zero spin density. It is known that gluons in QCD gain a dynamical mass
due to non-perturbative quantum corrections.
Appearance of the rest mass in classical spinless non-linear wave solutions 
implies possibility that corresponding spinless states may exist
in the quantum theory. The propagating non-linear waves described in Section II 
can be treated as a longitudinal massive modes in addition to the
Coleman transverse non-linear waves. It is clear, that
such of a set of the non-linear transverse and longitudinal plane waves
provides an attractive possibility for finding a proper non-perturbative quantization
of QCD. This raises also a question of finding a consistent definition
of the gluon spin density operator corresponding to the propagating spinless waves.
A strict concept of the gluon spin operator includes several aspects,
among them, the gauge invariance and Lorentz frame independence are the most important
issues. Below we construct a unique gauge invariant and Lorentz frame independent definition
of the gluon spin density operator for non-linear spinless waves at classical level.

Let us recall the main problems related to gauge invariant
definitions of gluon spin and angular momentum operators in QCD.
The standard canonical decomposition of the nucleon
total angular momentum includes gauge non-invariant
terms corresponding to quark and gluon spin and orbital
momentum operators \cite{JM}
\bea
J^{\rm can}_{\mu \nu}&=& \Int{\rm d}^3x\,
\Big\{ {\bar \psi}
\gamma^0 \dfrac{\Sigma_\mn}{2} \psi
-i  {\bar \psi} \gamma^0  x_{[\mu} \pro_{\nu]} \psi
-  A_{a[\mu}  F^a_{\nu] 0}
\nn \\
&&
\hspace*{12mm}
- F_{0 \rho}^a x_{[\mu} \pro_{\nu]} A^\rho_{a}\Big\}.
\quad \label{canon}
\eea
The expressions for the canonical spin density $S_{\mu\nu}=  -{A}_{a[\mu}  F^a_{\nu] 0}$ and angular momentum operators
for quark and gluon are gauge non-invariant.
It was assumed that a gauge invariant
definition of the gluon spin operator in non-Abelian
theory did not exist \cite{JM, ji}.
A novel approach towards constructing a gauge invariant
nucleon spin decomposition has been proposed in \cite{chen1,chen2}.
The main idea in this approach is to separate physical degrees of freedom
of gluon from pure gauge degrees of freedom, i.e., one splits
the gauge potential into two parts
\bea
A_\mu^a&=&{\cal A}_\mu^a+\hat A_\mu^a,
\eea
where
$\hat A_\mu^a$ is a physical gauge potential containing
only physical degrees of freedom, and
$ {\cal A}_\mu^a$ is a pure gauge potential satisfying the
pure gauge condition
\be
{\cal F}^a_{\mu\nu} \mbox{\small $({\cal A})$}=0. \label{puregauge}
%&& {\cal D}_i\vec A_i^{phys}=0, \label{condition2}
\ee
 Both potentials are constructed
in terms of the original unconstrained vector potential $A_\mu^a$. 
A key point is that the physical vector potential $\hat A_\mu^a$
transforms under the gauge transformation
as a vector in adjoint representation of $SU(2)$ whereas the pure gauge potential ${\cal A}_\mu^a$
transforms as a gauge connection.
This allows to re-write the canonical
decomposition (\ref{canon}) in an explicit gauge invariant manner \cite{chen1,chen2}
\bea
J^{\rm can}_{\mu \nu}&=& \Int{\rm d}^3x\,
\Big\{ {\bar \psi}
\gamma^0 \dfrac{\Sigma_\mn}{2} \psi
-i  {\bar \psi} \gamma^0  x_{[\mu} {\cal D}_{\nu]} \psi
-  \hat{A}_{a[\mu}  F^a_{\nu] 0}
\nn\\
&&\hspace*{13mm}
- F_{0 \rho}^a x_{[\mu} \big({\cal D}_{\nu]} \hat{A}^\rho_{a}
-{\cal F}_{\nu ]}{}^\rho{}_a({\cal A})\big)\Big\},
\label{canoncov}
\eea
where $ {\cal D}_\mu \equiv \pro_\mu+{\cal A}_\mu$ is a covariant derivative containing 
the pure gauge potential.
Each term in (\ref{canoncov}), including the gluon spin density operator
$\hat S_{\mu\nu}=-  \hat{A}_{a[\mu}  F^a_{\nu] 0}$,  is explicitly gauge invariant due to 
covariant transformation properties of the physical gauge potential $\hat{A}_\mu^a$.
It is worth to stress that an equation defining the physical
potential can be chosen by several ways, and the most
important issue in the definition of the physical potential
is a problem of uniqueness of gauge invariant and
Lorentz frame independent definition of spin and momentum
operators for quarks and gluon. Within the framework of the
formalism proposed by Chen et al \cite{chen1,chen2} a Lorentz frame independent 
definition is proposed in \cite{cho1},  however, an explicit construction
of such a definition is unknown since a perturbative solution
of the equation for the physical gauge potential does not exist on the space
of classical free plane wave solutions for the gluon field 
 (see also the review \cite{review1} for current status of the problem). 

Note that for massless particles a consistent concept of spin is given by the
notion of helicity. In the case of the Maxwell electrodynamics
the gauge potential must satisfy
two helicity gauge conditions,
\bea
A_0&=&0, ~~~~A_3=0, \label{hel1}
\eea
to represent helicity eigenstates of the
operator $J_3$ of a little group $E(2)$ which provides
naturally the Lorentz invariance of the helicity operator \cite{yskim}.
The helicity conditions can be expressed in equivalent forms
using various combinations of a generalized axial gauge condition
\bea
n^\mu A_\mu&=&0, \label{hel2}
\eea
where the constant vector $n^\mu$ specifies the temporal, axial or light-cone gauge
conditions.
%, and a Coulomb gauge type condition, $\pro^m A_m=0$.
The meaning of the helicity conditions is simple, they fix pure gauge degrees of freedom
while keeping two transverse dynamic degrees of freedom of the gluon.
Generalization of such description of the helicity operator to the case of
non-Abelian theory has been done in \cite{chen1, chen2,hatta,pakspin}.
 Since the helicity operator is Lorentz frame independent, all definitions satisfying the
helicity conditions  are consistent even though the
defining equations for the physical gauge potential
are not manifestly Lorentz invariant \cite{review1,pakspin}.

In the case of non-linear massive plane waves described by  the ansatz
(\ref{ansf1234}) the helicity conditions $A^a_0=0, ~ \pro^i A^a_i=0$ are fulfilled 
only in the rest frame, $\beta=0$. So that the definition of spin operator for such solutions
based on the helicity conditions becomes Lorentz frame dependent.
Solutions defined by the ansatz (\ref{wuyang},\ref{spherwav},\ref{spherwav3}) satisfy
the helicity conditions as well, $A_0^a=0, ~A_r^a=0$, but 
break the Lorenz gauge condition. In addition, one can verify that
the canonical spin density for all these solutions vanishes identically.
This implies that for massive spinless non-linear waves
one should find a proper definition for gluon spin operator. 
 
Note that the helicity conditions are consistent
with the Gauss law, which guarantees consistency with all equations of motion.
This is somewhat unexpected because the presence of helicity conditions
means that one has two transverse polarizations of gluon which
correspond to spin one particles. However, the non-linear wave type solutions
considered above admit projections of color fields along
the wave vector. This allows to interpret such non-linear waves as longitudinal modes,
raising a question about the number of dynamic degrees
of freedom for gluons.
It is clear, that the origin of such a subtlety comes from
the non-linearity of the gauge field strength expressed in terms
of the gauge potential. One possible way to revise
the notion of the gluon spin is to develope the old approach
based on constructing gauge invariant quantities from the field strength
\cite{weinberg64a,weinberg64b}.
Note that, so far the known gauge invariant definitions of the gluon spin operator
are conditioned by assumption that gluon has only two transverse dynamical
degrees of freedom per each color degree of freedom. Such an assumption is based on 
the standard perturbative quantization of QCD.
The presence of propagating massive wave solutions, which are essentially non-linear,
implies that definition of the gluon spin should be revised to include
description of zero spin states. 

Let us consider the definition for the physical gauge
potential based on the Lorenz gauge type equation
\be
\big({\cal D}^\mu \hat A_\mu\big)^a=0. \label{lorconstr}
\ee
One should stress that the last equation looks similar to Lorenz gauge fixing condition
in presence of a background field, however, it does not represent 
any gauge fixing in a fact. An explicit expression for the physical gauge potential $\hat A_\mu^a$ 
is provided as a solution to equation (\ref{lorconstr}) in terms of a general 
initial gauge potential $A_\mu^a$ in such a manner
that $\hat A_\mu^a$ still possesses full gauge freedom (it is transformed 
as $SU(2)$ vector in adjoint representation, not as a gauge connection). 
In practical use it is suitable to choose a real Lorenz gauge fixing condition 
for the physical potential with a trivial pure gauge counterpart, ${\cal A}_\mu^a=0$.
This is a key idea of the approach towards resolving the problem of
gauge invariant definitions for gluon spin and orbital momentum operators \cite{chen1, chen2}.

The equation (\ref{lorconstr}) is unique (except the case of a generalized covariant Lorenz gauge type
equation for the physical gauge potential which will be discussed below) among all possible conditions
containing first derivatives and satisfying invariance under the Poincare group transformations. 
The Lorenz gauge type condition (\ref{lorconstr})
was proposed in \cite{cho1} as a transversality condition.
In the case of Maxwell theory the definition of a gauge invariant photon spin operator
based on the solution of the Lorenz type equation (\ref{lorconstr}) encounters
a well-known problem of incompleteness of the Lorenz gauge \cite{pakspin}.
Namely, in the Lorenz gauge one has still a residual symmetry which
implies that equation of motion for the temporal component of the gauge potential
admits unphysical propagating modes. To fix such a symmetry one has to impose
an additional gauge condition.
We show that in the non-Abelian $SU(2)$ gauge theory the definition of a gauge invariant
spin operator based on the equation (\ref{lorconstr}) is possible 
due to absence of the residual symmetry for the class of
non-linear wave solutions described by ansatz (\ref{ansf1234},\ref{wuyang},\ref{spherwav},\ref{spherwav3}). 
To verify this we construct a formal series operator expansion
for such a solution using the perturbative method.
The solution for the physical potential $\hat A_\mu$ can be found
as a series
\be
\hat A^a_\mu=\hat A_\mu^{a(0)}+g \hat A_\mu^{a(1)}+g^2 \hat{A}_\mu^{a(2)}+
\cdots.
\ee
Expressing the pure gauge potential in terms of the
physical one, ${\cal A}^a_\mu=A^a_\mu-\hat A^a_\mu$,
one can find a solution to the equations (\ref{puregauge}) and (\ref{lorconstr})
in the leading order approximation
\bea
\hat A^a_\mu&=&
P_{\bot}{}_{\mu}{}^{\nu} A_\nu^a \nn \\
&& +g\dfrac{1}{\Box} P_{\bot}{}_\mu{}^\nu
\epsilon^{abc}\Big(
\pro^\rho(A_\rho^b A^c_\nu)
+\hat A_\rho^{b(0)} \pro^\rho \hat A_\nu^{c(0)} \nn \\
&& \hspace*{24mm}-\pro^\rho(A^b_\rho\hat A_\nu^{c(0)})
+\pro^\rho A_\nu^b\hat A_\rho^{c(0)}
\Big) \nn \\
&&
+\hat A_\mu^{a\, \rm long}, \label{solphys} \\
\hat A_\mu^{a\, \rm long}&=&
-g\epsilon^{abc}
\Big(\dfrac{1}{\Box}\pro^\rho A^b_\rho\Big)
\Big(P_{\bot}{}_{\mu}{}^\sigma A^c_\sigma\Big), \nn \\
\hat A_\mu^{(0)}&=& P_{\bot}{}_{\mu}{}^{\nu} A_\nu^a, \nn
\eea
where we use the transverse projectional operator
%$P_{=\mu\nu}=\frac{\pro_\mu \pro_\nu}{\Box}$ and
$P_\bot{}_\mu{}^\nu=\delta_\mu^\nu-\frac{\pro_\mu \pro^\nu}{\Box}$.
Note that the solution (\ref{solphys}) for the physical gauge potential
includes an expression $\hat A_\mu^{a\, \rm long}$ containing the longitudinal
part,  $\pro^\rho A_\rho^a$, of the unconstrained gauge potential. 
The fact that a gauge invariant expression (\ref{solphys})
is given in terms of the gauge potential, not in terms of the field strength,
reflects the property of the non-Abelian gauge theory
where the field strength does not determine all gauge invariant quantities
as it occurs in the Abelian gauge theory.
One should stress that non-locality appearing in the solution for
the physical gauge potential is unphysical. Such a non-locality disappears
when we impose the real Lorenz gauge fixing condition on the physical potential;
in this case the pure gauge potential vanishes identically and
the physical gauge potential coincides with the initial general
gauge potential $A_\mu^a$. So that the non-local expression for the
gluon spin density operator reduces to the standard local expression
for the canonical spin density with $A_\mu^a$ satisfying the usual Lorenz gauge condition. 

Let us verify that the Lorenz gauge condition is complete
and does not possess a residual symmetry on a space of the
non-linear wave type solutions. A small gauge variation of the
Lorenz gauge condition leads to the following equation
for the gauge parameter $\lambda^a$
\bea
(\pro^\mu D_\mu \lambda)^a=0. \label{lorcond3}
\eea
For simplicity we consider a class of plane wave solutions provided by
the ansatz  (\ref{ansf1234}).
One can verify that any solution to equation (\ref{lorcond3})  for
the parameters $\lambda^a$
does not belong to the same class of the non-linear
plane wave solutions determined by the ansatz (\ref{ansf1234}).

Let us consider a counter example, a generalized Lorenz gauge
$\big(D^\mu A_\mu\big)^a=0$
which contains a full covariant derivative $D^\mu$ including
the gauge potential.
Under small gauge variation it
leads to a covariant D'Alembert equation
\bea
(D^\mu D_\mu \lambda)^a=0.   \label{eqlambda}
\eea
Applying the ansatz  (\ref{ansf1234})
for homogeneous solutions, one obtains
the following system of equations
\bea
&& \ddot\lambda_1+g\lambda_1 (\phi_2^2+\phi_3^2)=0, \nn \\
&& \ddot\lambda_2+g\lambda_2 (\phi_3^2+\phi_1^2)=0,  \\
&& \ddot\lambda_3+g\lambda_3 (\phi_3^2+\phi_1^2)=0. \nn
\eea
It is clear, that the above system of equations on the space of
the non-linear plane wave solutions $(\ref{phi0})$ admits solutions for $\lambda^a$
exactly of the same type. This implies appearance of a
residual symmetry at the level of a full nonlinear theory
and makes the covariant Lorenz gauge
incomplete. So that, a defining equation for the physical gauge potential
based on use of a generalized covariant Lorenz gauge type equation can not
provide a self-consistent definition of a gauge invariant and Lorentz covariant gluon spin operator.

Summarizing, we propose a new definition for the gluon spin operator 
for a class of non-linear plane wave solutions which admit a non-vanishing mass and a total spin zero.
The definition for the physical gauge potential is based on the equation (\ref{lorconstr})
and we  have provided an explicit construction for the physical gauge potential, (\ref{solphys}),
which leads to unique gauge invariant and Lorentz frame independent definition of the gluon spin operator 
for massive spinless non-linear plane waves. Note, that equation (\ref{lorconstr}) was suggested 
in \cite{cho1} as a possible definition of the gluon spin operator for massless gluons
with helicity $\pm1$. However, the equation (\ref{lorconstr}) does not admit solutions 
due to presence of the residual symmetry. So that, a consistent
definition of the gluon spin operator for massless gluons 
is possible only on the basis of the helicity equations (\ref{hel1}, \ref{hel2}).

\section{Conclusion}

We propose a new stationary generalized monopole solution which
can be treated (at least in the singular Abelian gauge) 
as a system of the static Wu-Yang monopole and off-diagonal gluon.
An essential feature of that solution is 
that it possesses a finite energy density everywhere in the whole space.

For a class of propagating non-linear waves described by the ansatz (\ref{ansf1234})
we construct a unique gauge invariant and Lorentz frame independent definition
for the gluon spin operator based on a Lorenz gauge type equation for
the physical gauge potential.
Note that in the case of the stationary monopole solution
defined by the ansatz (\ref{spherwav})  the canonical spin density
vanishes identically as well. However, the Lorenz gauge condition
is not fulfilled
\bea
\pro^\mu A_\mu^a&=&\delta^{a,2} \dfrac{1}{r^2}\cot \theta \psi(r,t).
\eea
This is not surprising, since the stationary monopole solution does not represent
a stationary propagating solution like a plane wave. Passing to arbitrary Lorentz frame
such a solution will represent  a moving lump with a maximal energy density around its center mass.
Due to this, the definition of the gluon spin operator for the stationary
generalized Wu-Yang monopole solutions and for non-stationary solutions corresponding to the ansatz (\ref{spherwav})
is based on a generalized axial gauge condition  as for the massless gluon \cite{pakspin}.
One should note that, since the notion of spin
has a quantum mechanical origin, one has to perform properly the quantization 
of the gluon field and construct the algebra of quantized
angular momentum and spin operators. Since for the case of non-linear solutions the
angular momentum and spin operators are non-linear 
operator functions of the quantized gauge potential $A_\mu^a$,
one expects that such a quantum algebra can be realized as
a deformation or a non-linear representation of the standard Lie algebra of $SO(3)$ .

An important issue is to study possible physical implications of non-linear wave type solutions
in QCD. It has been shown that non-linear type I plane waves, (\ref{phi0}), provide
a simple estimate of glueball spectrum in qualitative agreement
with lattice calculation \cite{frasca09}. 
Our stability analysis shows that the stationary spherically symmetric monopole is unstable under axially-symmetric perturbations. This indicates to existence of stationary monopole and monopole-antimonopole pair solutions. This issue will be considered in a subsequent paper  \cite{p2}.

\acknowledgments
%{\bf Acknowledgements}
One of authors (DGP) acknowledges Prof. Mo-Lin Ge and Prof. C.M. Bai
for warm hospitality during his staying in Chern Institute of Mathematics,
and Dr. E. Tsoy for numerous discussions.
The work is supported by NRF of Korea, grant MSIP No.2014R1A2A1A01002306(ERND), 
and by Brain Pool Program (KOFST), Sogang University; 
by NSFC (Grants 11035006 and 11175215),
Rare Isotope Science Project of
Inst. for Basic Sci. funded by Ministry of Science, ICT and Future
Planning and NRF of Korea (2013M7A1A1075764), and by UzFFR (Grant F2-FA-F116).

%\clearpage
%\vspace{2mm}

\end{document}